\documentclass[preprint,groupedaddress]{revtex4}
\usepackage[dvips]{graphicx}
\usepackage{epsfig}
\usepackage{graphicx}
\def\be{\begin{equation}}
\def\ee{\end{equation}}
\def\ba{\begin{eqnarray}}
\def\ea{\end{eqnarray}}

\begin{document}
\title{Nuclear-induced time evolution of entanglement of two-electron spins in anisotropically coupled quantum dot}
\author{Gehad Sadiek,$^{1,2,3}$ Zhen Huang,$^{4}$ Omar Aldossary,$^{1}$ and Sabre Kais$^{4}$}
\address{$^1$Department of Physics, King Saud University, Riyadh 11451-2455, Saudi Arabia}
\address{$^2$Department of Physics, Purdue University, West Lafayette, IN 47907, USA}
\address{$^3$Department of Physics, Ain Shams University, Cairo 11566, Egypt}
\address{$^4$Department of Chemistry and Birck Nanotechnology Center, Purdue University, West Lafayette IN 47907, USA}
\begin{abstract}
We study the time evolution of entanglement of two spins in anisotropically coupled quantum dot interacting with the unpolarized nuclear spins environment. We assume that the exchange coupling strength in the z-direction $J_z$ is different from the lateral one $J_l$. We observe that the entanglement decays as a result of the coupling to the nuclear environment and reaches a saturation value, which depends on the value of the exchange interaction difference $J=\| J_l-J_z\|$ between the two spins and the strength of the applied external magnetic field. We find that the entanglement exhibits a critical behavior controlled by the competition between the exchange interaction $J$ and the external magnetic field. The entanglement shows a quasi-symmetric behavior above and below a critical value of the exchange interaction. It becomes more symmetric as the external magnetic field increases. The entanglement reaches a large saturation value, close to unity, when the exchange interaction is far above or below its critical value and a small one as it closely approaches the critical value. Furthermore, we find that the decay rate profile of entanglement is linear when the exchange interaction is much higher or lower than the critical value but converts to a power law and finally to a Gaussian as the critical value is approached from both directions. The dynamics of entanglement is found to be independent of the exchange interaction for isotropically coupled quantum dot.
\end{abstract}
\maketitle
\section{Introduction}
Coherence and entanglement lie in the heart of quantum mechanics since its birth and are of fundamental interest in modern physics. Particular fields where they play a crucial role are quantum information processing and quantum computing \cite{Nielsen, Bouwmeester, gruska, macchiavelleo}. Decoherence is considered as one of the main obstacles toward achieving a practical quantum computing system as it acts to randomize the relative phases of the two-state quantum computing units (qubits) due to coupling to the environment\cite{Decoherence}. Quantum entanglement is a nonlocal correlation between two (or more) quantum systems such that the description of their states has to be done with reference to each other even if they are spatially well separated. Entanglement arises naturally when manipulating linear superpositions of quantum states to implement the different proposed quantum computing algorithms \cite{Shor, Grover}. Different physical systems have been proposed as reliable candidates for the underlying technology of quantum computing and quantum information processing \cite{Barenco, ibm-stanford, NMR1, NMR3, TrappedIones, CvityQED, JosephsonJunction}. There has been special interest in solid state systems as they can be utilized to build up integrated networks that can perform quantum computing algorithms at large scale. Particularly, the semiconductor quantum dot is considered as one of the most promising candidates for playing the role of a qubit \cite{spin-qgate, spin-qubit, spin-orbit}. The main idea is to use the spin {\bf S} of the valence electron on a single quantum dot as a two-state quantum system, which gives rise to a well-defined qubit. On the other hand, the strong coupling of the electron spin on the quantum dot to its environment stands as a real challenge toward achieving the high coherent control over the spin required for information and computational processing. The main mechanisms responsible for spin decoherence are spin-orbit coupling and spin-nuclear spins (of the surrounding lattice) coupling \cite{spin-orbit, spin-nucspin}. Recent experiments show that the spin relaxation time due to spin-orbit coupling (tens of ms) is about six orders of magnitude longer than that due to coupling to nuclear spins (few ns) 
\cite{Kroutfar, Elzerman, Florescu}. This made the electron spin coupling to the nuclear spins the target of large 
number of theoretical \cite{theory-Golovach, theory-Coish, theory-Shenvi, theory-Klauser, theory-analytical} and 
experimental \cite{experim-Huttel, experim-Tyryshkin, experim-Abe, experim-Johnson, experim-Koppens, experim-Petta} research works. The dipolar interaction between the nuclei in 
the dot does not conserve the total nuclear spin and as a result the change in the nuclear spin configuration happens within $T_{nuc} \sim 10^{-4} s$ which is the precession time of a single nuclear spin in the local magnetic field of the other spins. This precession time is much longer than the decoherence time of the electron spin due to hyperfine coupling in the dot ($\sim 10^{-6}$ s), therefore we can safely ignore this interaction as well as the fluctuation of the nuclear magnetic field when treating the entanglement and decoherence problem of the electron spins. 
In addition to the proposals for using the single quantum dot as a qubit, there has been others for using coupled quantum dots as quantum gates\cite{spin-qgate, spin-orbit}. The aim is to find a controllable mechanism for forming an entanglement between the two quantum dots (two qubits) in such a way to produce a fundamental quantum computing gate such as XOR. In addition, we need to be able to coherently manipulate such an entangled state to provide an efficient computational process. Such coherent manipulation of entangled states has been observed in other systems such as isolated trapped ions \cite{trapped-ions} and superconducting junctions \cite{supercond-junc}. Recently an increasing effort to investigate entangled states in coupled quantum dots has emerged. The coherent control of a two-electron spin state in a coupled quantum dot was achieved experimentally, where the coupling mechanism is the Heisenberg exchange interaction between the electron spins \cite{experim-Johnson, experim-Koppens, experim-Petta}. The mixing of the two-electron singlet and triplet states due to coupling to the nuclear spins was observed. The induced decoherence  (leading to the singlet triplet mixing) in the system has been studied theoretically, where each one of the two electrons were assumed to localized on its own dot and coupled to a different nuclear spin environement \cite{theory-coher-control-1, theory-coher-control-2, huang1}. There has been proposals for experimental scheme to create, detect and control entangled spin states in coupled quantum dots \cite{proposed-scheme}. Recently, there has been an increasing interest in studying the anisotropy in the exchange interaction in coupled quantum dot systems, which is mainly due to the spin-orbit coupling, the unavoidable asymmetry of the dots structure and the effect of the external magnetic field \cite{Anisotropic_J1,Anisotropic_J2,Anisotropic_J3,Anisotropic_J4}. Furthermore, there has been proposals to utilize this anisotropy, rather than removing it, in introducing a set of quantum logic gates to be implemented in quantum computing \cite{Anisotropic_J_QComput1,Anisotropic_J_QComput2,Anisotropic_J_QComput3}.

In this paper we study the dynamics of entanglement of two electron spins
confined in a system of two laterally coupled quantum dots with one net
electron per dot. The two electron spins are coupled through exchange
interaction. Modeling the coupling between the two electrons on a couple dot system by exchange interaction, including the low barrier case, has been studied in detail in Ref. \cite{spin-orbit}. We assume that the exchange coupling is anisotropic such that the two spins couple in the transverse (z-direction) with a coupling strength that is different from the one in the lateral direction (x and y-directions). An external magnetic field is applied perpendicularly to the plane of confinement. We assume that the potential barrier between the two dots is low enough that the two electrons are delocalized over the two dots and as a result are coupled to a common nuclear environment which splits the two spins space into two unmixable subspaces (one contains the singlet state and the other contains the three triplets). Therefore, the singlet-triplet mixing can not take place for this system set up in contrast to the case studied in Ref. \cite{theory-coher-control-1} and \cite{theory-coher-control-2}. Nevertheless, mixing within the triplets subspace is possible and in order to study it we assume that the system was initially prepared in a maximally entangled triplet spin state with a total antisymmetric wave function, which is experimentally an achievable task \cite{experim-Koppens, experim-Petta}. The two spins are coupled through hyperfine interaction to the surrounding unpolarized nuclear spins in the two
dots. We study the time evolution of the entanglement of the two spins induced by the nuclear spins at different strengths of the external magnetic field and exchange interaction, which can be tuned and measured experimentally by controlling the different gate voltages on the two dots \cite{experim-Petta}. We found that 
the entanglement which is initially maximum decays as a result of coupling to the nuclear spins within a decoherence time of the order of $10^{-4}$. The entanglement decay rate and its saturation value vary over wide range and are determined by the value of the exchange interaction difference (between the transverse and lateral components) and the strength of the external magnetic field. The entanglement exhibits a critical behavior which is decided based on the competition between the exchange interaction difference (regardless of its sign) and the external magnetic field, it shows a quasi-symmetric behavior above and below a critical value of the exchange interaction difference which is determined based on the value of the external magnetic field. The behavior becomes more symmetric as the external magnetic field increases. Comparable decay rates and saturation values were observed for the entanglement on both sides of the critical exchange interaction value. Far from the critical exchange value, the entanglement shows high saturation values and slow power law decay rates. Close to the critical value, the saturation values become much lower and the decay rate shows Gaussian profiles. Our results suggest that for maintaining the entanglement between the two spins, the difference between the values of the exchange interaction difference and the external magnetic field should be tuned to be large. This paper is organized as follows. In sec. II we introduce our model and the calculations of the spin correlation functions. In sec. III we show the evaluation of entanglement. In sec. IV we present our results, discussing and explaining the different important findings we obtained. We close with our conclusions in Sec V.
\section{Two electron spins coupled to a bath of nuclear spins}
We consider two coupled quantum dots with a net single valence electron per
dot. The two electron spins $\bf{S_1}$ and $\bf{S_2}$ couple to each other
through exchange interaction and couple to the same external magnetic field $\bf{B}$ and to the common unpolarized nuclear spins on the two dots $\{ \bf{I}_i\}$ through hyperfine interaction. The exchange interaction is taken to be anisotropic with the coupling strength in the z-direction to be $J_z$ and in the x and y directions as $J_l$.
The spin-orbit coupling and the dipolar interactions between the nuclei are ignored. The system is described by the Hamiltonian
\be
\hat{H} = J_z \; S_{1z} S_{2z}  +  J_l \; \left( S_{1x} S_{2x}+ S_{1y} S_{2y} \right) + g \mu_B \; \bf{S} \cdot B +  \bf{S} \cdot \bf{h}_{2N},
\ee
where g is the electron $g$ factor and $\mu_B$ the Bohr magneton. $\bf{S} = \bf{S_{1}} + \bf{S_{2}}$ and $ \bf{h}_{2N}$ $ = \sum_{i} A_i \bf{I}^i = g \mu_B \bf{H}_{2N}$, where $\bf{H}_{2N}$ is the two dot nuclear magnetic field and the sum is running over the entire space. $A_i$ is the coupling constant of $\bf{S}$ with the $\it{i}$th nucleus and is equal to $A v_0 |\psi_A (\bf{r_i})|^2$, where A is a hyperfine constant, $v_0 (= a^2 a_z / N) $ is the volume of the crystal cell, $a$ and $a_z$ are the single dot sizes in the lateral and transverse directions, and $\psi_A (\bf{r}_i)$ is the two electrons envelope antisymmetric wave function at the nuclear site $\bf{r}_i$. For a typical single quantum dot size, the number of nuclei $N = 10^6$ and their typical nuclear magnetic field affecting the electron spin through hyperfine coupling is of the order of $\sim A/(\sqrt{N}g \mu_B)$ \cite{optical-orientation}, which is approximately of magnitude $\simeq 5$ mT \cite{experim-Petta}, with an associated electron precession frequency $\omega_N \simeq A/\sqrt{N}$, where $\omega_N \gg 1/T_{nuc}$, and $a \sim 15$ nm.
The presence of an external magnetic field causes Zeeman splitting for the electron spin and in addition has a polarizing effect on the nuclear spins. The magnitude of these effects and the associated electron precession frequency depends on the strength of the applied field which we consider as a varying parameter in this work and is changed over a wide range. Therefore, the precession frequency about the external magnetic field can be higher or lower than $\omega_N$ depending on the strength of the external field compared to the nuclear field. 
We consider some particular unpolarized nuclear configuration, represented in terms of the $\hat{I}_{iz}$ nuclear spin eigenbasis as $|{I_{z}^{i}}\rangle$, with $I_{i}^{z}=\pm 1/2$, where we have considered nuclear spin 1/2 for simplicity. This nuclear configurations have the same time evolution as the more general initial tensor product states corresponding to arbitrary directions of individual spins \cite{spin-nucspin}.
The Hamiltonian $H$ (Eq. 1) is symmetric under exchange of the two electron spins $\bf{S_1}$ and $\bf{S_2}$ which reflects the fact that in this model the two electrons are considered indistinguishable (delocalized over the two dots). This Hamiltonian splits the the two spins space into two subspaces, one of them contains the singlet state and the other contains the three triplet states. In this system set up (described by the Hamiltonian $H$), the two subspaces never get mixed. The only mechanism by which these two subspaces can get mixed is that each electron spin couples to a different nuclear magnetic field (or even a different magnetic field), which is not the case here.
We assume that the two coupled electron spins are initially $(t=0)$ in a triplet, maximally entangled, state described in the coupled representation by $|T_0\rangle = |\Uparrow\Downarrow + \Downarrow\Uparrow\rangle$. 
We are interested in investigating the dynamics of entanglement of this state (dynamics of entanglement and its non-monoticity behavior has been studied in many works, for example see Ref. [40] and references therein). For that purpose we evaluate the correlators 
\be 
C_{ij}(t)=\langle n| \hat S_{i}(t) \hat S_{j}(t) |n \rangle, \hspace{2cm} i,j=x,y,z
\ee
where $|n \rangle$ is the total system (two electrons and nuclei) state given by $|T_0\rangle |\{I_{iz}\}\rangle$ and $\hat S_{i}(t)= e^{it\hat{H}} \hat S_{i} e^{-it\hat{H}}$. The initial state $|n \rangle$ is an eigenstate of the Hamiltonian 
\be
\hat{H}_0 = J_z \hat{S}_{1z} \hat{S}_{2z} + \frac{1}{2}J_l(\hat{S}_{1+}\hat{S}_{2-}+\hat{S}_{1-}\hat{S}_{2+}) 
+ \epsilon_z \sum_{i=1}^2 \hat{S}_{iz} 
+\hat{h}_{2Nz} \sum_{i=1}^2 \hat{S}_{iz},
\ee
with an eigenenergy $\epsilon_n = -J_z/4 + J_l/2$, where $\epsilon_z = g \mu_B B_z$, consequently we can expand in the perturbation (where we use the fact that for a typical nuclear configuration $ A_{k}^2 / h_{2n} << 1$ which acts as our expansion parameter as can be noticed in the coming calculations \cite{spin-nucspin, optical-orientation})
\be
\hat{V}= \sum_{i=1}^2(\hat{S}_{i+} \hat{h}_{2N-}+\hat{S}_{i-}\hat{h}_{2N+}),
\ee
where the total Hamiltonian $\hat{H}=\hat{H_0}+\hat{V}$. Using the time evolution operator $\hat{U}(t)= \hat{T} \exp[-it\int_{0}^{t}dt' \hat{V}(t')]$, where $\hat{T}$ is the usual time-ordering operator, we obtain
\be
C_{ij}(t) = C_{ij}^{(0)} + C_{ij}(t)
\ee
where $C_{ij}^{(0)}$ is the correlator corresponding to $\hat{V} = 0$, while $C_{ij}(t)$ is corresponding to the corrections due to the perturbation $\hat{V}$ and are given by
\be
C_{ij}^{(0)} = \langle n| \hat S_{i} \hat{S}_{j}|n \rangle
\ee
and
\ba
C_{ij}(t) &=& \langle n|\hat{U}^{\dagger}(t) \hat{S}_{i}(t) \hat{U}(t) \hat{U}^{\dagger}(t) \hat{S}_{j}(t) \hat{U}|n \rangle
\nonumber\\
&=& \sum_{k_1} \langle n|\hat{U}^{\dagger}(t) \hat{S}_{i}(t) \hat{U}(t) |k_1\rangle
\langle k_1| \hat{U}^{\dagger}(t) \hat{S}_{j}(t) \hat{U}|n \rangle 
\nonumber\\
&+&
\sum_{k_2} \langle n|\hat{U}^{\dagger}(t) \hat{S}_{i}(t) \hat{U}(t) |k_2\rangle
\langle k_2| \hat{U}^{\dagger}(t) \hat{S}_{j}(t) \hat{U}|n \rangle
\ea
where the intermediate state $|k_1> = |T_{+1}\rangle |\{I_{z}^{i}\}\rangle=| \Uparrow\Uparrow \rangle |\{\cdot\cdot\cdot ,I_{z}^{k_{1}}= -1/2, \cdot\cdot\cdot \}\rangle$ and $|k_2> = |T_{-1}\rangle |\{I_{z}^{i}\}\rangle=| \Downarrow\Downarrow\rangle |\{\cdot\cdot\cdot ,I_{z}^{k_{2}}= +1/2, \cdot\cdot\cdot \}\rangle$, while for the singlet state $|S_{0}\rangle$ we have $\langle T_{0}|\hat{V}|S_{0}\rangle = 0$. The non-vanishing correlators to the leading order in $\hat{V}$ read
\be
S^z_{12}=\langle n| \hat S_{1z} \hat S_{2z} |n \rangle = -\frac{1}{4} +\frac{1}{4} (\Gamma_1(t) + \Gamma_2(t)) \:,
\ee
\be
S^x_{12}=\langle n| \hat S_{1x} \hat S_{2x} |n \rangle = \frac{1}{4} \:,
\ee
\be
S^y_{12}=\langle n| \hat S_{1y} \hat S_{2y} |n \rangle = \frac{1}{4} \:,
\ee
\be
\label{M1z}
M^z_1=\langle n| \hat S_{1z}|n \rangle = \frac{1}{2} (\Gamma_1(t) - \Gamma_2(t)) \:,
\ee
\be
\label{M2z}
M^z_2=\langle n| \hat S_{2z}|n \rangle = \frac{1}{2} (\Gamma_1(t) - \Gamma_2(t)) \:,
\ee
where
\be
\label{first-order}
\Gamma_i(t) = 4 \sum_{k_i} \frac{|V_{n,k_i}|^2}{\omega_{n,k_i}^2} \sin^{2}(\omega_{n,k_i} t/2) \:, \hspace{3 cm}  i= 1, 2
\ee
where $V_{n,k_1}= \langle n |\hat{V}| k_1\rangle = A_{k_1}/\sqrt{2}$ is the matrix element between initial state $|n\rangle = |T_0 \rangle |\{\cdot\cdot\cdot ,I_{z}^{k_{1}}= 1/2, \cdot\cdot\cdot \} \rangle$ and the intermediate state $|k_1\rangle$ and $V_{n,k_2} = A_{k_2}/\sqrt{2}$ is between  $|n\rangle = |T_0 \rangle |\{\cdot\cdot\cdot ,I_{z}^{k_{2}}= - 1/2, \cdot\cdot\cdot \} \rangle$ and $|k_2\rangle$ and 
\be
\label{omega_k1}
\omega_{n,k_1} = \epsilon_n - \epsilon_{k_1} = [\frac{J}{2}-(\epsilon_{z}+(h_z)_{2n})]- \frac{1}{2}A_{k_1}\equiv \omega_{-} - \frac{1}{2}A_{k_1} \:,
\ee
\be
\label{omega_k2}
\omega_{n,k_2} = \epsilon_n - \epsilon_{k_2} = [\frac{J}{2}+(\epsilon_{z}+(h_z)_{2n})]+ \frac{1}{2}A_{k_2} \equiv \omega_{+} + \frac{1}{2}A_{k_2} \:,
\ee
where $J=J_l-J_z$ is the exchange interactions difference due to the anisotropy of the coupling and $(h_z)_{2n}=\langle n|h_{2Nz} |n \rangle \simeq \langle n|\sum_{i \ne k_i} I_{z}^i|n \rangle$. We will use $h_{2n}$ instead of $(h_z)_{2n}$ from now on for simplicity.
As a result of the large value of N, we can replace the sum over $k_i$ by integrals ($\sum_{k_i} F_{k_i} = 1/v_0 \int F({\bf r}) + O(1/N)$) and we obtain
\be
\Gamma_1(t) = \frac{1}{\omega_{-}^2}[I_0 -I_1 \cos (\omega_{-} \; t) - I_2 \sin(\omega_{-} \; t)] \:,
\ee
\be
\Gamma_2(t) = \frac{1}{\omega_{+}^2}[I_0 -I_1 \cos (\omega_{+} \; t) + I_2 \sin(\omega_{+} \; t)] \:,
\ee
where
\be
I_{0}= \frac{1}{v_0} \int dx dy dz [\mathcal A (x,y,z) t]^2\:,
\ee
\be
I_{1}= \frac{1}{v_0} \int dx dy dz  [\mathcal A (x,y,z)]^2 \cos[\mathcal A (x,y,z) t / 2]\:,
\ee
\be
I_{2}= \frac{1}{v_0} \int dx dy dz  [\mathcal A (x,y,z)]^2 \sin[\mathcal A (x,y,z) t / 2]\:,
\ee
and where 
\be
{\mathcal A}(x,y,z) = A v_{0} |\psi(x,y,z)|^2 \:.
\ee
$|\psi(x,y,z)|^2$ is the two electrons envelope antisymmetric wave-function which we construct starting from the two single state wave functions  (which is the typical wave function used to describe the electron state in the confining potential of a quantum dot system \cite{spin-orbit, spin-nucspin})
\ba
\phi_1(x_1,y_1,z_1) = e^{-(x_{1}-b)^2/a^2} e^{-y_{1}^2/a^2} e^{-z_{1}^2/a_{z}^2},    
\nonumber\\          
\phi_2(x_2,y_2,z_2) = e^{-(x_{2}+b)^2/a^2} e^{-y_{2}^2/a^2} e^{-z_{2}^2/a_{z}^2} 
\ea
which are mixed to obtain an antisymmetric wave function $\psi(x_1, y_1, z_1; x_2, y_2, z_2)$. Finally integrating out the coordinates of one of the two electrons, squaring and normalizing we obtain
\be
|\psi(x,y,z)|^2 =  \frac{\sqrt{2}}{\pi^{3/2} a^2 a_z \sinh(2b^2/a^2)} e^{-2 y^2/a^2} e^{-2 z^2/a_{z}^2} e^{-2 x^2/a^2} [\cosh(4 b x/a^2) - e^{-2 b^2/a^2}]\:,
\ee
where $2b$ is the interdot distance. In our calculations we set $b=a/2$. The integrals $I_0$, $I_1$ and $I_2$ were evaluated numerically.
\section{Evaluating the entanglement of formation}
\indent In this paper we investigate the time evolution of two-spin entangled
state by calculating the entanglement of formation $E$ between them. The concept of entanglement of formation is related to the amount of entanglement needed to obtain the maximum entanglement state $\rho$, where~$\rho$~is the density matrix in the two-spin uncoupled representation basis. It was shown by Wootters\cite{Wootters98} that 
\begin{equation}
E(\rho)=\mathcal{E}(C(\rho)),
\end{equation}
\noindent where the function $\mathcal{E}$ is given by
\begin{equation}
\mathcal{E}=h(\frac{1+\sqrt{1-C^2}}{2}),
\end{equation}
where $h(x)=-xlog_2x-(1-x)log_2(1-x)$ and the 
concurrence $\textit{C}$ is defined as
\begin{equation}
C(\rho)=max\{0,~\lambda_1-\lambda_2-\lambda_3-\lambda_4~\}.
\end{equation}
 For a general state of two qubits, 
 $\lambda_i$'s are the eigenvalues, in decreasing order, of the 
 Hermitian matrix
\begin{equation}
R \equiv~\sqrt{\sqrt{\rho}~\tilde{\rho}~\sqrt{\rho}},
\end{equation}
\noindent where ~$\tilde{\rho}$~ is the spin-flipped state of the density matrix $\rho$, and defined as\\
\begin{equation}
\tilde{\rho}=(\sigma_y \otimes \sigma_y) \rho^*(\sigma_y \otimes \sigma_y).
\end{equation}
where $\rho^*$ is the complex conjugate of $\rho$.

Alternatively, the $\lambda_i$'s are the square roots of the eigenvalues 
of the non-Hermitian $\rho \tilde{\rho}$. Since the density matrix $\rho$ 
follows from the symmetry properties of the Hamiltonian, the $\rho$ must be real and symmetrical\cite{Osterloh02}, plus
the symmetric property in X and Y direction in our system ,we obtain 
\begin{equation}
\rho={\left(\begin {array}{cccc} \rho_{1,1} 
& 0 & 0 & 0 \\ 
0 & \rho_{2,2} & \rho_{2,3} & 0 \\
 0 & \rho_{2,3} & \rho_{3,3} & 0\\ 
0 & 0 & 0 & \rho_{4,4}\end {array} \right)}.\\
\end{equation}

\noindent The $\lambda_i$'s can be written as

\begin{equation}
\label{eigenvalue}
\lambda_a = \lambda_b= \sqrt{\rho_{1,1} \rho_{4,4}},~ 
\lambda_c= |\sqrt{\rho_{2,2} \rho_{3,3}} + | \rho_{2,3} ||,~
\lambda_d =| \sqrt{\rho_{2,2} \rho_{3,3}} - | \rho_{2,3} ||.
\end{equation}

Using the definition $<A>=Tr(\rho A)$, we can express all the 
matrix elements in the density matrix in terms of the different spin-spin correlation functions \cite{huang2, huang3}:
\begin{center}
\begin{equation}
\rho_{1,1}=\frac{1}{2} M_1^z + \frac{1}{2} M_2^z + S_{12}^z + \frac{1}{4},
\end{equation}
\begin{equation}
\rho_{2,2}=\frac{1}{2} M_1^z -\frac{1}{2} M_2^z - S_{12}^z + \frac{1}{4},
\end{equation}
\begin{equation}
\rho_{3,3}=-\frac{1}{2} M_1^z +\frac{1}{2} M_2^z - S_{12}^z + \frac{1}{4},
\end{equation}
\begin{equation}
\rho_{4,4}= - \frac{1}{2} M_1^z -\frac{1}{2} M_2^z + S_{12}^z + \frac{1}{4},
\end{equation}
\begin{equation}
\rho_{2,3}= S_{12}^x + S_{12}^y,
\end{equation}
\end{center}

Note that if the sign of $J$ is inverted (i.e. $J \rightarrow -J$), $\omega_{+} \rightarrow -\omega_{-}$ and vice versa and as a result $\Gamma_1$ and $\Gamma_2$ exchange their expressions ($\Gamma_1 \leftrightarrow \Gamma_2$). The only physical quantities that are affected by this sign flip are $M_{1}^{z}$ and $M_{2}^{z}$ (Eqs. (\ref{M1z}), (\ref{M2z})) which flip sign as well. Despite these sign flips the eigenvalues of the matrix $R$ given by Eq. \ref{eigenvalue} are not affected because under this sign change the density matrix elements $\rho_{1,1} \leftrightarrow \rho_{4,4}$ and $\rho_{2,2} \leftrightarrow \rho_{3,3}$ whereas the other elements are invariant.
\section{results and discussions}
\begin{figure}
\begin{center}
\includegraphics[width=1.0\textwidth,height=0.6\textheight]{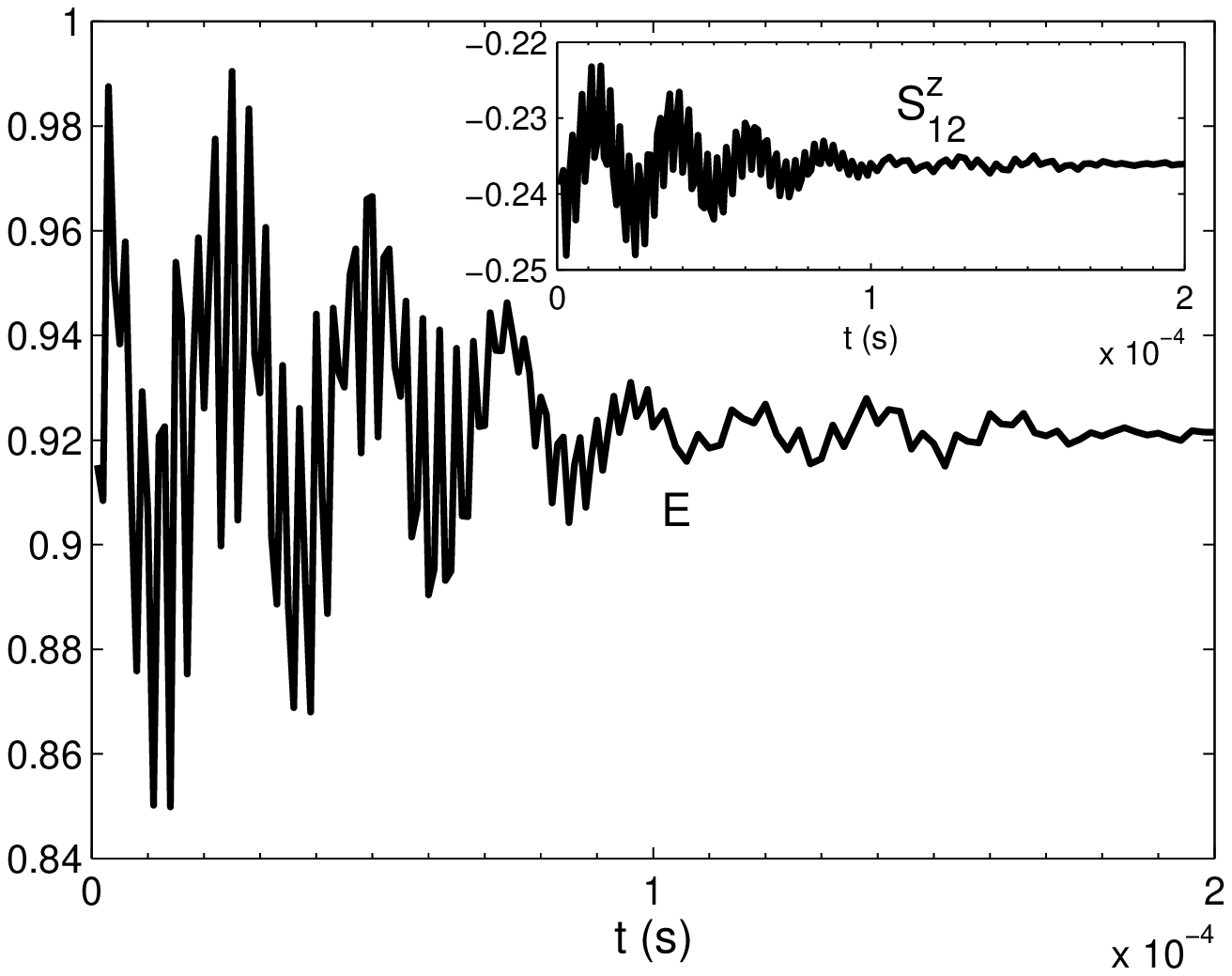}
\end{center}
\caption{Time evolution of entanglement and spin correlation in z-direction in absence of external
magnetic field and for very small anisotropy $J=A/5000$. The nuclear magnetic field $h_{2n}=A/500$, where A is the hyperfine constant.}
\label{fig1}
\end{figure}

\begin{figure}
\begin{center}
\includegraphics[width=1.0\textwidth,height=0.6\textheight]{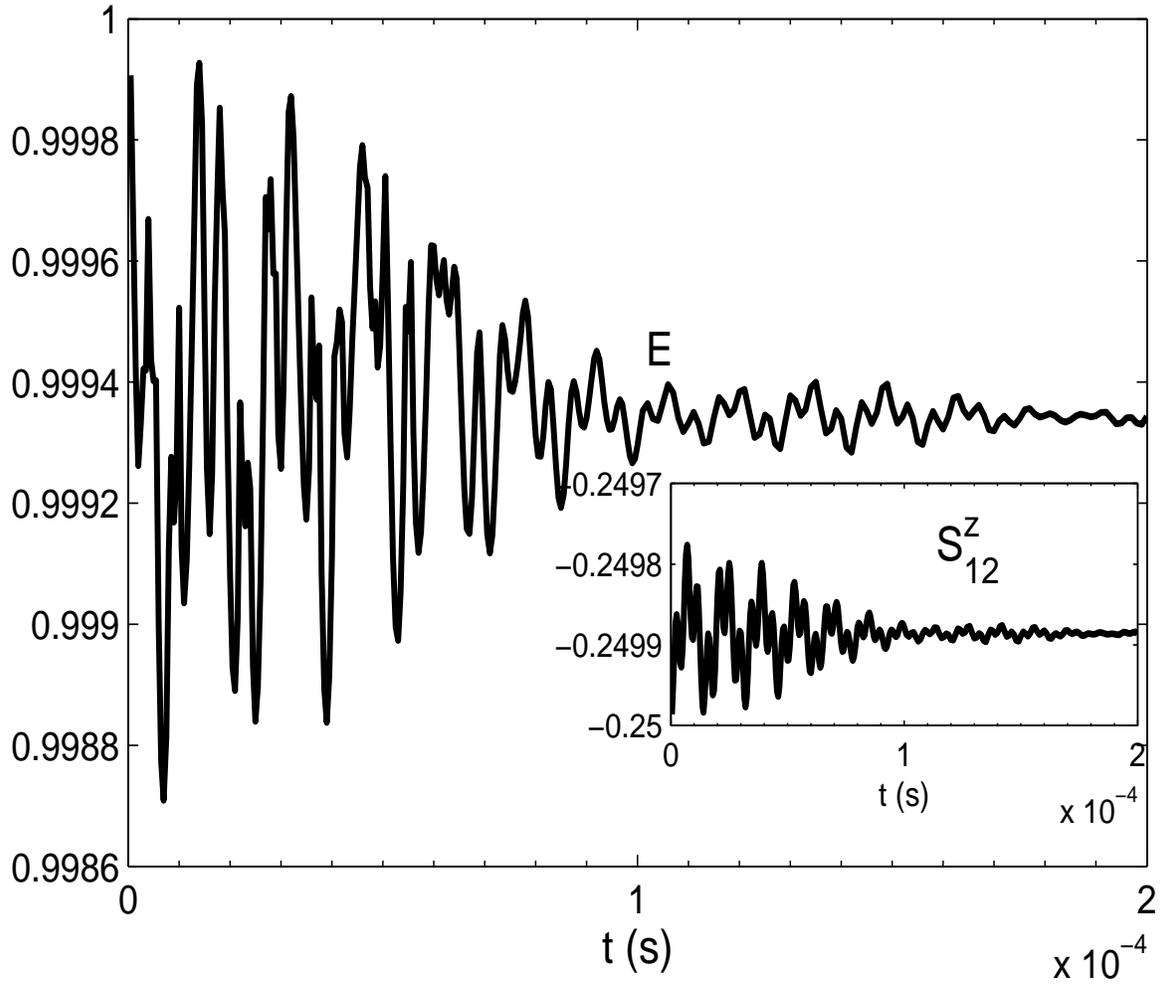}
\end{center}
\caption{Time evolution of entanglement and spin correlation in the z-direction for very small anisotropy $J=A/5000$ with a large external magnetic field $\epsilon_z=A/50 >> h_{2n}$.}
\label{fig2}
\end{figure}

\begin{figure}
\begin{center}
\includegraphics[width=1.0\textwidth,height=0.6\textheight]{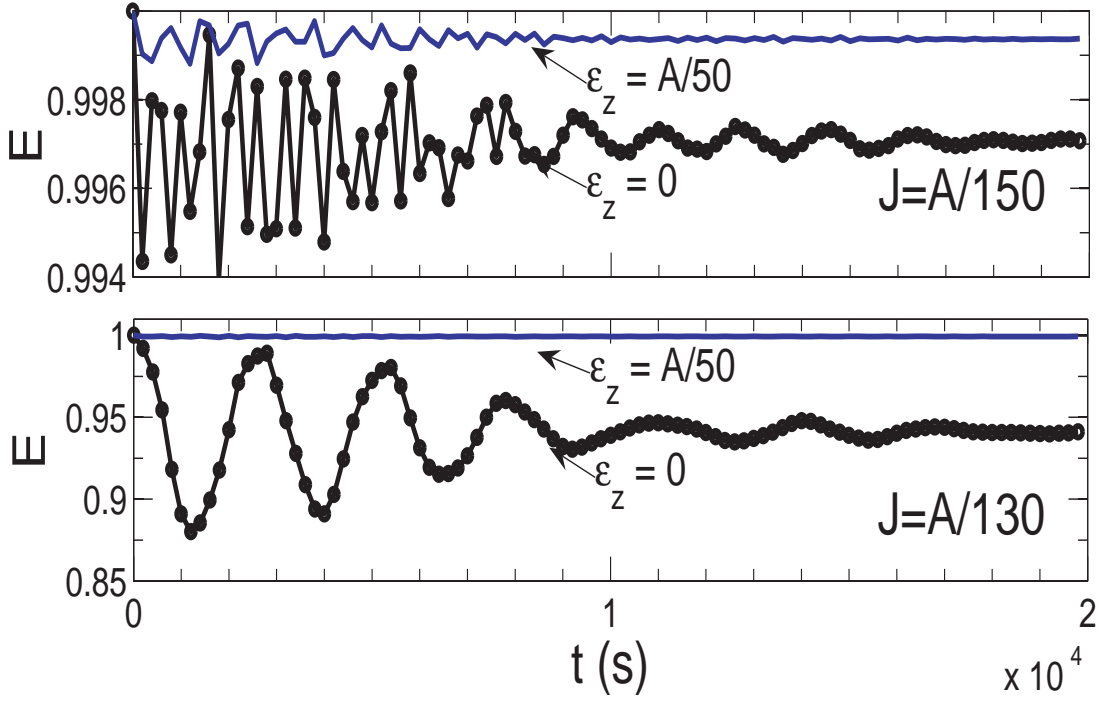}
\end{center}
\caption{Dynamics of entanglement at different exchange interaction differences $A/150,$ and $A/130$ with the external magnetic field $\epsilon_z$ set at 0 and $A/50$.}
\label{fig3}
\end{figure}

\begin{figure}
\begin{center}
\includegraphics[width=1.0\textwidth,height=0.6\textheight]{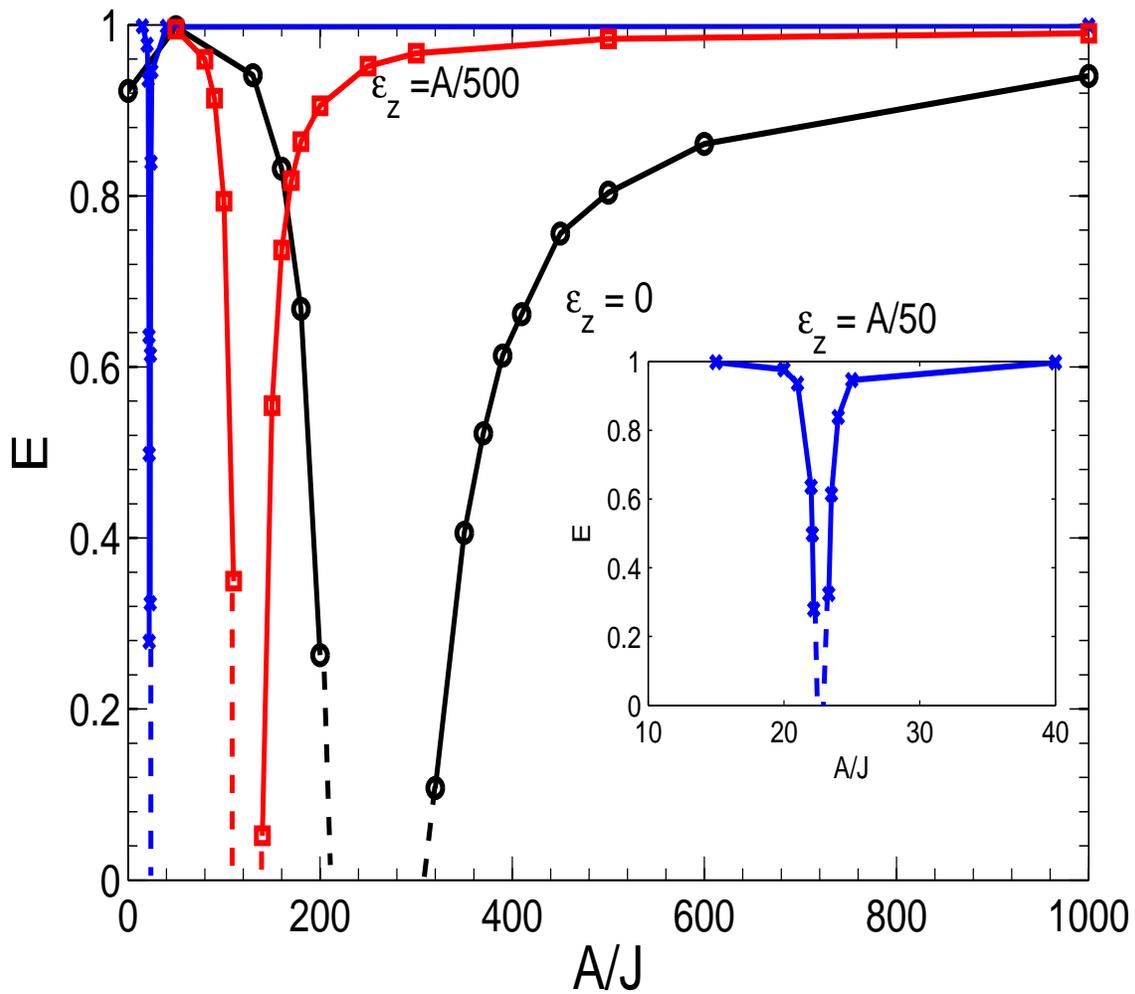}
\end{center}
\caption{The phase diagram of entanglement. The saturation value of the entanglement is plotted vs. the exchange interaction difference J at different values of the external magnetic field $\epsilon_z=0$, $A/500$, and $A/50$. The case corresponding to $\epsilon_z=A/50$ is replotted in the inner panel to demonstrate the symmetric behavior of E at high external magnetic field.}
\label{fig4}
\end{figure}

In this paper, we focus on the dynamic behavior of entanglement of two coupled electron spins under different external magnetic field strengths and at different values of the Heisenberg exchange interactions difference $J$. Interestingly, the entanglement is independent of the sign of the exchange interaction difference $J$, as was explained in sec. 3, which means that regardless of which coupling component is dominant ($J_l$ or $J_z$) the behavior of the entanglement as a function of $J$ stays the same.
Firstly, we study the nuclear-induced time evolution of entanglement in absence of external magnetic field and with a very small anisotropy in the exchange interaction. In fig. \ref {fig1}, we plot the entanglement of two quantum dots versus time with $J=A/5000$ and $\epsilon_z=0$. The initial state of the coupled quantum dot is triplet {\bf $\frac{1}{\sqrt{2}}(|\Uparrow \Downarrow \rangle + |\Downarrow \Uparrow \rangle)$}, where the two spins are maximally entangled. As shown in fig. \ref{fig1}, the entanglement is initially maximum (E=1). Once the interaction between the nuclear magnetic field and quantum dots is turned on, the entanglement starts decaying with a rapid  oscillation and reaches a saturation value of about 0.921 within a decay time of the order of $10^{-4}$s. As can be noticed the envelope of this oscillating entanglement decreases with a power law profile under this condition before reaching saturation. From the process of constructing the density matrix to calculate the entanglement, the entanglement, magnetization and spin correlations are directly related. We plot the time evolution of the spin correlation function in the z-direction $S_{12}^{z}$ in the inner panel. The correlation function exhibits a very similar behavior to that of the entanglement, it begins with a maximum value, -0.25, then decays to a saturation value of about -0.236 within the same decay time. Under the first order perturbation theory, the spin correlations in x and y directions are unaffected by nuclear spins. This leads to the higher saturation value of entanglement in our calculation, which may change at higher orders. Furthermore, because of the initial unpolarized state for the nuclear spins and absence of external magnetic field, the magnetization of the two quantum dots remains zero. 
The effect of the external magnetic field with a very small anisotropy in exchange coupling between the two spins is shown in fig. \ref{fig2} where $\epsilon_z$ is set to be $A/50>>h_{2n}$ and $J=A/5000$. The entanglement evolves with a similar decaying behavior to what we have observed in fig. \ref{fig1} except that the decaying amplitude of oscillation is much smaller and the saturation value, for both entanglement and correlation, is much higher (0.99935 and -0.24989 respectively), which emphasis the important role the external magnetic field plays, as expected, significantly enhancing the entanglement between the two spins.
In fig. \ref{fig3}, we focus on the role played by exchange interaction difference $J$. The dynamics of entanglement is considered at different values of $J$ at two different specific external magnetic fields, zero and $A/50$. In the two panels, $J$ is set to be $A/150$ and $A/130$ in the upper and lower panels respectively. Interestingly, comparing with fig. 1, the saturation value of the entanglement increases, from 0.92 to 0.997, as we increase $J$ from $A/5000$ to $A/150$ but then decreases back, to 0.93, as we increase $J$ further to $A/130$. This indicates that the larger $J$ does not always help maintaining the entanglement between the two spins, i.e. the entanglement is not a monotonic function of J. Obviously, there is an inversion point of E between these values of J. The effect of the magnetic field is clear in the three case, it suppresses the amplitude of the decaying oscillation and enhances the entanglement as a result of its polarizing effect, giving raise to a higher saturation value in the three cases. \\
In fact, the entanglement shows very similar behavior at different magnitudes of the external magnetic field and exchange interaction difference, however the asymptotic saturation values and decay rates vary significantly based on the values of J and $\epsilon_z$. In principle, each electron spin is in a precession motion about two fields, magnetic which is uniform and the non-uniform nuclear field. This precession motion is the reason behind the oscillatory behavior of the entanglement, whereas coupling to the non-uniform nuclear field is responsible for the decay where the spin energy is dissipated to the nuclear environment (the same behavior was discussed in detail in Ref. [18] for a single spin on a single dot). Therefore, it is of great interest to study the mutual effect of the exchange interaction difference and the external magnetic field on the entanglement. The phase diagram of the system is shown in fig. \ref{fig4} where The saturation value of entanglement is plotted versus $J$ at different values of the external magnetic field. As the entanglement mostly reaches its saturation value at $t \sim 10^{-4} s$, we use the value of E at $t=10^{-4}$ to approximately represent the saturation value at each J. There are three pairs of curves with each pair corresponding to a specific value of the external magnetic field given by, from left to right, $A/50, A/500, 0$. The pair of curves corresponding to $\epsilon_z=A/50$ are replotted in the inner panel for clarity. 
It is very interesting to note that for each $\epsilon_z$ the behavior of the entanglement is almost symmetric about a critical value of J, we call it $J_c$. On each pair, the entanglement saturation value is initially unity or very close to unity ,depending on the external magnetic field strength, then starts to decrease as J increases. It decays exponentially as $J$ approaches $J_c$ but once $J$ exceeds $J_c$ it raises up exponentially again reaching a value of unity or close. The decay rate of the entanglement saturation value for $J > J_c$ is more rapid than $J < J_c$ but they become more symmetric as $\epsilon_z$ increases. Surprisingly, this phase diagram shows a very peculiar role played by the exchange interaction difference, namely that small values $(J<<J_c)$ and great values as well $(J >> J_c)$ enhances the entanglement between the two spins, while as $J \rightarrow J_c$ it is reduced significantly. 

\begin{figure}
\begin{center}
\includegraphics[width=1.0\textwidth,height=0.6\textheight]{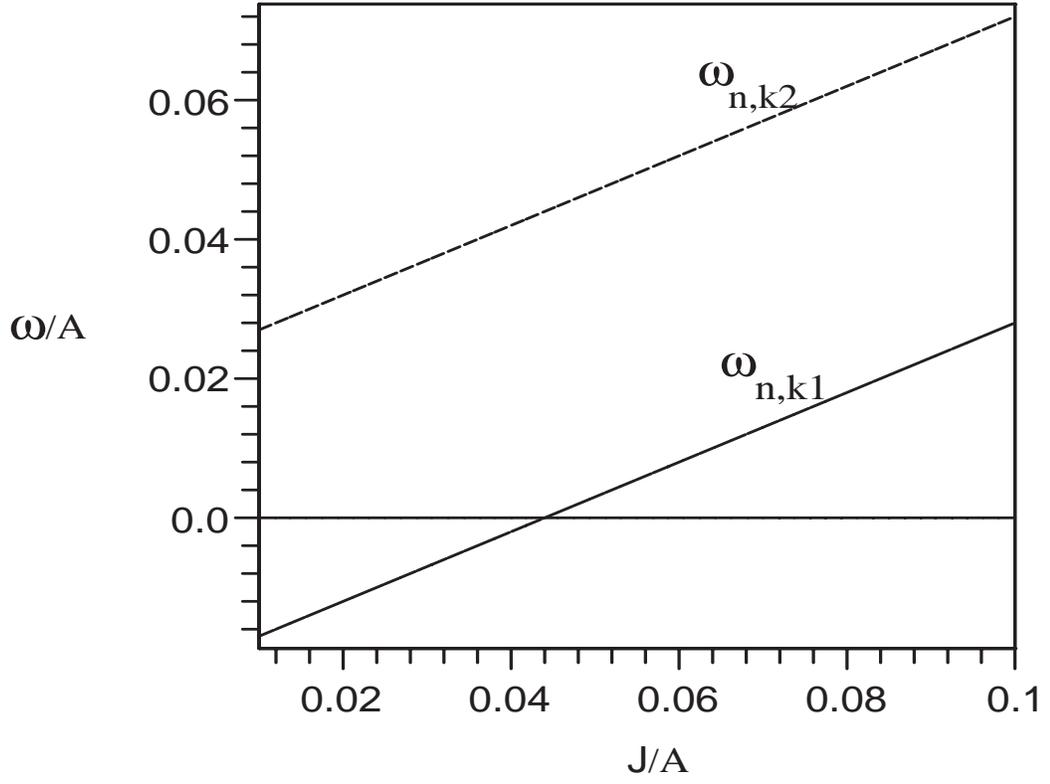}
\end{center}
\caption{The energy differences $\omega_{n,k1}$ (solid line) and $\omega_{n,k2}$ (dashed line) are plotted versus the exchange interaction difference J for chosen $h_{2n}=A/500$ and $\epsilon_z=A/50$.}
\label{fig5}
\end{figure}

Unfortunately, we can not investigate the entanglement behavior when $J$ is very close or equal to $J_c$ because the system diverges under this condition. The reason for this divergence is the degeneracy (electronic levels crossing) that takes place between the initial state and the intermediate states which causes a break down of the first-order perturbation theory. This degeneracy is manifested in fig.~\ref{fig5}, where the energy difference $\omega_{n,k1}$ between the initial state $|n\rangle$ and the first intermediate state $|k_1\rangle$ and the difference $\omega_{n,k2}$ between $|n\rangle$ and $|k_2\rangle$ (given by Eqs.~\ref{omega_k1} and~\ref{omega_k2}) are plotted versus the exchange interaction difference $J$ for chosen $h_{2n}=A/500$ and $\epsilon_z=A/50$. As can be seen in fig.~\ref{fig5}, $\omega_{n,k1}$ vanishes at a particular value of $J$ which turns out to be the same as $J_c$ observed in the inner panel in fig.~\ref{fig4} where the same values of parameters are used in that case. Since $\omega_{n,k1}$ appears squared in the denominator of the first order perturbation term (Eq.~\ref{first-order}), it is responsible for the divergence when it vanishes. Also, the symmetric behavior of the entanglement about $J_c$ is due to the linear dependence of the magnitude of $\omega_{n,k1}$ on $J$ as it increases or decreases away from $J_c$. Similarly, $\omega_{n,k2}$ may cause the same effect if negative values of $J$ were considered.
The value of the critical exchange interaction $J_c$ corresponding to each $\epsilon_z$ can be easily deduced from the phase diagram to be equal to $2(h_{2n}+\epsilon_z)$ coinciding with the observations of fig.~\ref{fig5}, which emphasis that the external magnetic field does not only enhance the entanglement but also dictates the value of $J_c$.

\begin{figure}
\begin{center}
\includegraphics[width=1.0\textwidth,height=0.6\textheight]{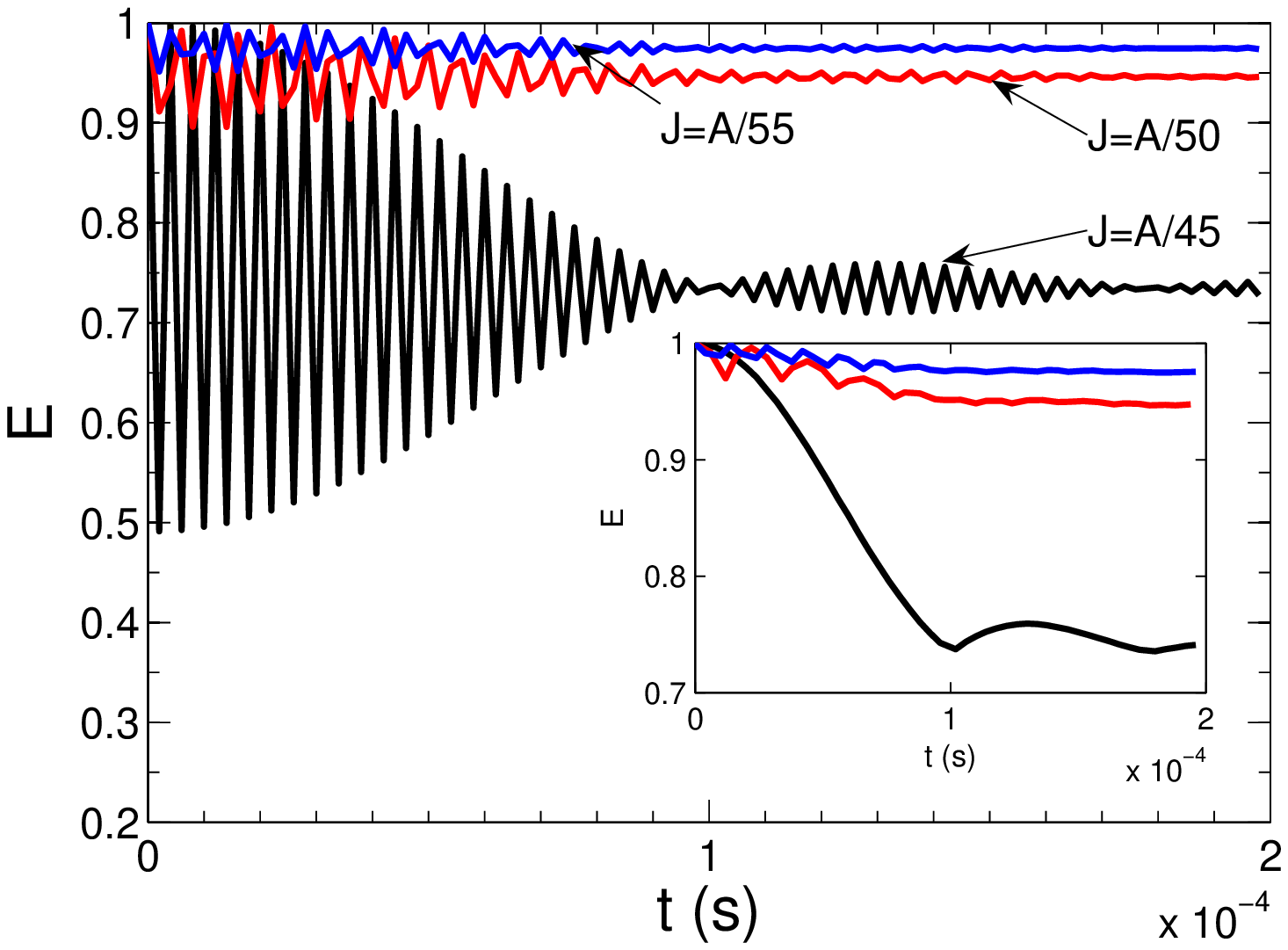}
\end{center}
\caption{Dynamics of entanglement with the external magnetic field fixed at $\epsilon_z=A/100$
while the exchange interaction difference $J$ is set at values below the critical value ($J_c=A/41.7$), $J=A/45$, $A/50$ and $A/55$. The inner panel shows the corresponding entanglement decay profiles.}
\label{fig6}
\end{figure}

\begin{figure}
\begin{center}
\includegraphics[width=1.0\textwidth,height=0.6\textheight]{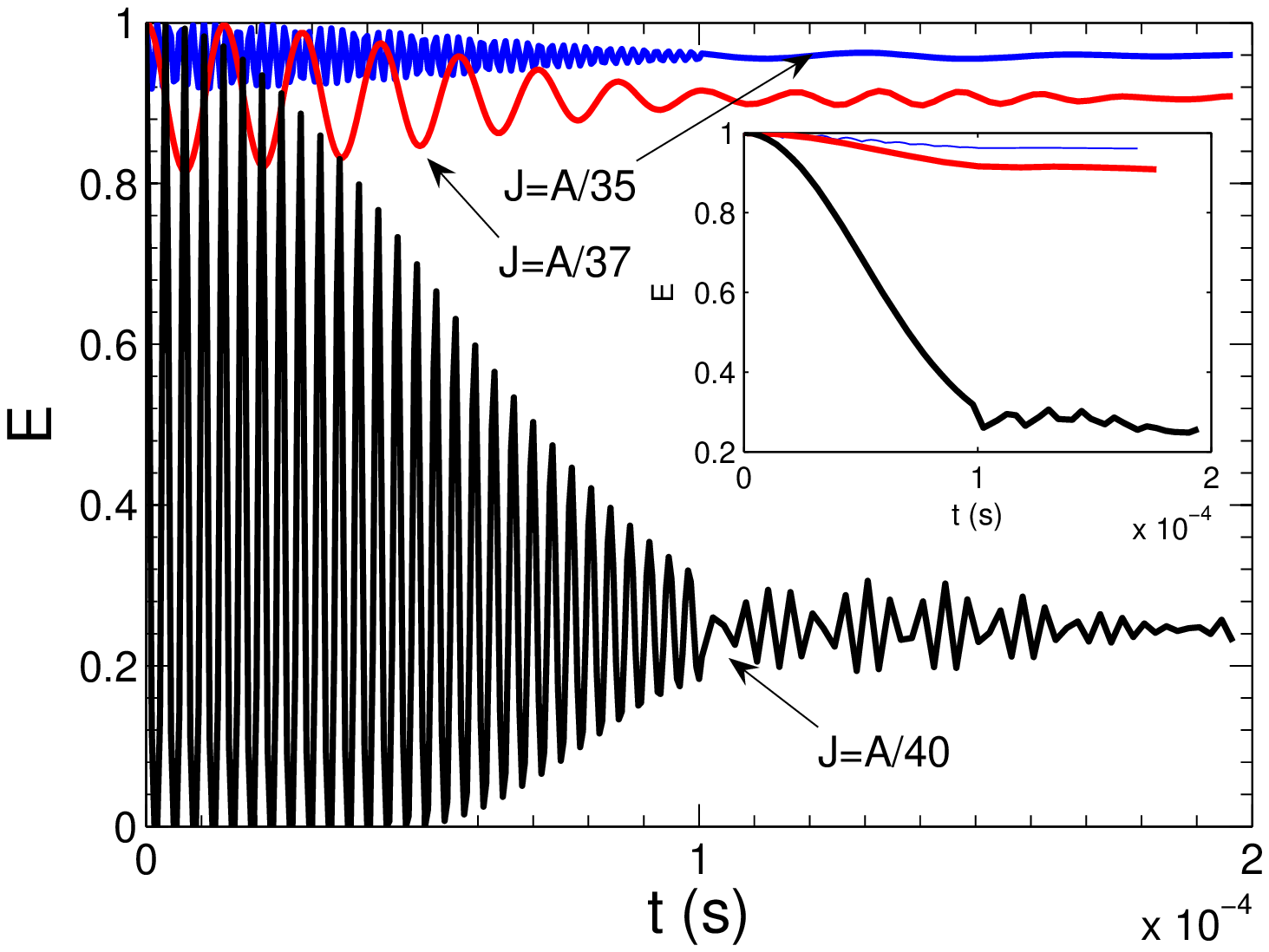}
\end{center}
\caption{Dynamics of entanglement with the external magnetic field fixed at $\epsilon_z=A/100$
while the exchange interaction difference $J$ is set at values above the critical value ($J_c=A/41.7$), $J=A/35$, $A/37$ and $A/40$. The inner panel shows the corresponding entanglement decay profiles.}
\label{fig7}
\end{figure}

After the important observations from the phase diagram (fig. \ref{fig4}), in particular concerning the entanglement critical behavior upon varying J, it is interesting to further investigate the effect of $J$ on the entanglement above, below and close to $J_c$ at a fixed value of the external magnetic field. Therefore, the dynamics of entanglement at different exchange interaction values below $J_c$ in fig. \ref{fig6} and above it in fig. \ref{fig7} is examined. Where $\epsilon_z$ is set to be $A/100$ and consequently $J_c$ turns out to be $A/41.7$. In fig. \ref{fig6} we consider the exchange interaction values $A/45, A/50$ and $A/55$ $(< J_c)$ whereas in fig. \ref{fig7} $A/40, A/37$ and $A/35$ $(> J_c)$. 
As we can see in the two graphs, when the value of $J$ is far from $J_c$ , either higher or lower, the entanglement oscillation is small and the decay rate is slow reaching high saturation values that approach unity as J becomes much smaller or much higher than $J_c$. As J gets very close to $J_c$ the entanglement exhibits very rapid oscillation with very large initial amplitude and decays very rapidly as well reaching a low saturation value, 0.74, for $J < J_c$ and very low one, 0.22, for $J > J_c$. The contrast between those values gets smaller and smaller with increasing the magnetic field as can be noticed from the phase diagram (fig. \ref{fig4}). To compare the decay rates of entanglement at different $J$ values, the corresponding entanglement decay profiles are plotted in the inner panels in figs. \ref{fig6} and \ref{fig7}. As shown, the decay profiles are linear when $J$ is much higher (or lower) than $J_c$ converting to a power-law and then to a Gaussian as J gets closer to $J_c$ from above or below. The critical behavior of the entanglement based on the strength of $J$ points out that there is a competition between the effects of the external magnetic field, which enhances the entanglement through polarizing the two spins in its direction, and the exchange interaction difference, which enhances the entanglement on the other hand by favoring anti-ferromagnetic correlation. Therefore when one of the two effects is dominant, i.e. $J<< \epsilon_z$ (or $>> \epsilon_z$), there will be a large net correlation one way or the other leading to strong entanglement. Tuning the two effects to be comparable reduces the net correlation significantly and forces the entanglement to attenuate to a low value.
\section{conclusions}
In Summary, we study the time evolution of entanglement of two electron spins in anisotropically coupled quantum dot under coupling with nuclear spins in an applied external magnetic field in the triplet $({\bf S} = 1)$ subspace. Our main observations is that the dynamics of entanglement exhibits a critical behavior determined by the competition between the value of the exchange interaction difference between the two spins and the strength of the external applied magnetic field. When one of them dominates, the entanglement saturates to a very large value close to the maximum. On the other hand, when they are comparable the entanglement attenuates and may reach very low saturation values. We found that for each fixed value of the external magnetic field there is a corresponding critical value of the exchange interaction difference where the entanglement between the two spins as a function of the exchange interaction exhibits symmetrical behavior about it. When we studied the time evolution of the entanglement at values of the exchange interaction difference above and below the critical value, we observed that the entanglement decay profiles are linear when the exchange interaction value is much higher or lower than the critical value and saturates to large values but then converts to a power law and finally to a Gaussian as it approaches the critical value from above or below reaching low saturation values. The system diverges when the exchange interaction value becomes very close or equal to the critical value which might be a sign of a critical behavior taking place in the environment consisting of the nuclear spins. For an isotropic coupling the entanglement becomes absolutely independent of the Heisenberg exchange interaction. Our observations suggest that tunning the applied external magnetic field
and (or) the exchange interaction difference between the two spins in such a way that they differ significantly in value may sustain the correlation between the two spins in the coupled quantum dot and lead to a strong entanglement between them.
\section*{Acknowledgments}
This work has been supported in part by King Saud University, College of Science Research Center, grant no. Phys2006/19. One of the authors (G.S.) would like to thank E. I. Lashin for useful discussions.

\newpage

\end {document}